\documentclass[12pt]{interact}

\usepackage{epstopdf}
\usepackage{subfigure}

\usepackage{amsmath,bm,amssymb}
\usepackage{graphicx,xcolor,soul}
\usepackage{tabularx,ragged2e,booktabs}
\usepackage[colorlinks = true, linkcolor = blue, urlcolor  = cyan, citecolor = red, anchorcolor = blue]{hyperref}

\usepackage[numbers,sort&compress,merge]{natbib}

\begin{document}

\title{Laser-Induced Electron Diffraction: Inversion of Photoelectron Spectra for Molecular Orbital Imaging}

\author{
\name{
	R. Puthumpally-Joseph\textsuperscript{1,2}\thanks{Contact: raijumon.puthumpally-joseph@u-psud.fr},
	J. Viau-Trudel\textsuperscript{1,3},
	M. Peters\textsuperscript{4},
	T. T. Nguyen-Dang\textsuperscript{3},
	O.\;Atabek\textsuperscript{1} and
	E. Charron\textsuperscript{1}}
\affil{
	\textsuperscript{1}Institut des Sciences Mol\'eculaires d'Orsay (ISMO), Univ. Paris-Sud, CNRS, Universit\'e Paris-Saclay, 91405 Orsay cedex, France\newline
	\textsuperscript{2}Laboratoire Interdisciplinaire Carnot de Bourgogne (ICB), UMR 6303 CNRS-Universit\'e Bourgogne Franche Comt\'e, 9 Av. A. Savary, BP 47 870, F-21078 Dijon cedex, France\newline
	\textsuperscript{3}D\'epartement de Chimie, Universit\'e Laval, Qu\'ebec, Canada G1K 7P4\newline
	\textsuperscript{4}Universit\'e de Moncton, Edmundston, NB, Canada E3V 2S8}
}

\maketitle

\begin{pacscode}
33.80.-b, 34.80.Qb, 34.80.Bm, 42.50.Hz
\end{pacscode}

\begin{abstract}
In this paper, we discuss the possibility of imaging molecular orbitals from photoelectron spectra obtained via Laser Induced Electron Diffraction (LIED) in linear molecules. This is an extension of our work published recently in Physical Review A \textbf{94}, 023421 (2016) to the case of the HOMO-1 orbital of the carbon dioxide molecule. We show that such an imaging technique has the potential to image molecular orbitals at different internuclear distances in a sub-femtosecond time scale and with a resolution of a fraction of an Angstr\"om.
\end{abstract}


\section{Introduction}

Unlike in the linear regime of light-matter interaction, the response of an atom or a molecule to an intense external field depends profoundly on the field parameters. Following the theoretical predictions of L. V. Keldysh in 1965\;\cite{JETP.20.1307}, significant developments in laser technologies have enabled researchers to observe tunnel ionization using strong infrared (IR) laser fields. When an intense IR field is applied to an atom or a molecule it distorts the system by forming a potential barrier through which a bound electron can tunnel out. The subsequent dynamics of the laser-driven electron wave packet in the continuum can be described by the so-called three-step mechanism introduced by P. B. Corkum in 1993\;\cite{PRL.71.1994}. After its release in the continuum, the electron, following the field, will scatter from the ionic core. This process is known as recollision. The particular case of elastic recollision is popularly known as \emph{Laser Induced Electron Diffraction}, or LIED\;\cite{CPL.259.313}. This strong field effect and its acronym were first thought of by A. D. Bandrauk and coworkers. In 1996, T. Zuo, A. D. Bandrauk and P. B. Corkum suggested that this laser-induced diffraction process could be used to study the dynamics of molecular systems and to access structural information about molecules \cite{CPL.259.313}. These predictions were first demonstrated experimentally in 2008\;\cite{Sc.320.1478}. Since then, several theoretical and experimental LIED related studies were performed in order to image time-resolved dynamics of molecular systems\;\cite{PRA.83.051403,PRA.85.053417,Nat.483.194,NatCom.6.7262,JPB.49.112001}. Closely related to LIED is the attosecond photoelectron holography imaging idea, again pioneered by A. D. Bandrauk and coworkers\;\cite{PRL.108.263003, PRA.84.043420}. 

A  breakthrough was achieved for imaging molecular orbitals in 2004 using a tomographic technique based on the measurement of high harmonics (HHG) generated by inelastic recollision events\;\cite{Nat.432.867,Nat.Phys.7.822}. HHG based tomographic techniques are relatively complex but they provide accurate results for the reconstruction of molecular orbitals. We have shown recently that LIED also has the potential to be developed into such an imaging tool. Indeed, drawing the connection with optical diffraction, LIED photoelectron spectra can be seen as an image of the scattering centers in the reciprocal space. Inverting the LIED spectrum of a molecule back to real space, should thus allow to obtain an image of the scattering centers, and therefore of the molecule itself.

We have demonstrated that it is possible to invert the diffraction patterns of LIED spectra obtained from the Highest Occupied Molecular Orbital (HOMO) of CO$_{2}\/$\;\cite{PRA.in.press}. In the HOMO, only two oxygen atoms of the molecule  contribute to the ionization signal. The inversion problem is more involved when all atoms of the molecule  contribute. In this paper, we present such a case where all the atoms of a linear molecule have significant contributions, using the HOMO-1 orbital of CO$_{2}\/$ as an example.

The paper is organized as follows: In section \ref{sec:theo} we describe the theoretical models used in this study. This section presents both the numerical and analytical approaches we have developed. Section \ref{sec:reconst} discusses the extraction of information about the molecule from the LIED spectra of the HOMO-1 orbital of CO$_{2}\/$. The reconstruction procedure is discussed in detail in this section. Finally in section \ref{sec:conclude}, we conclude our work. Note that atomic units are used throughout the paper unless stated otherwise.

\section{Theoretical models}\label{sec:theo}

The aim of this paper is to extend the inverse problem for imaging molecular orbitals we discussed in Ref.\;\cite{PRA.in.press} to the cases where the initial state has a different symmetry and shape compared to the HOMO of CO$_2$.

\begin{figure}[ht]
	\centering
	\includegraphics[width=0.95\textwidth]{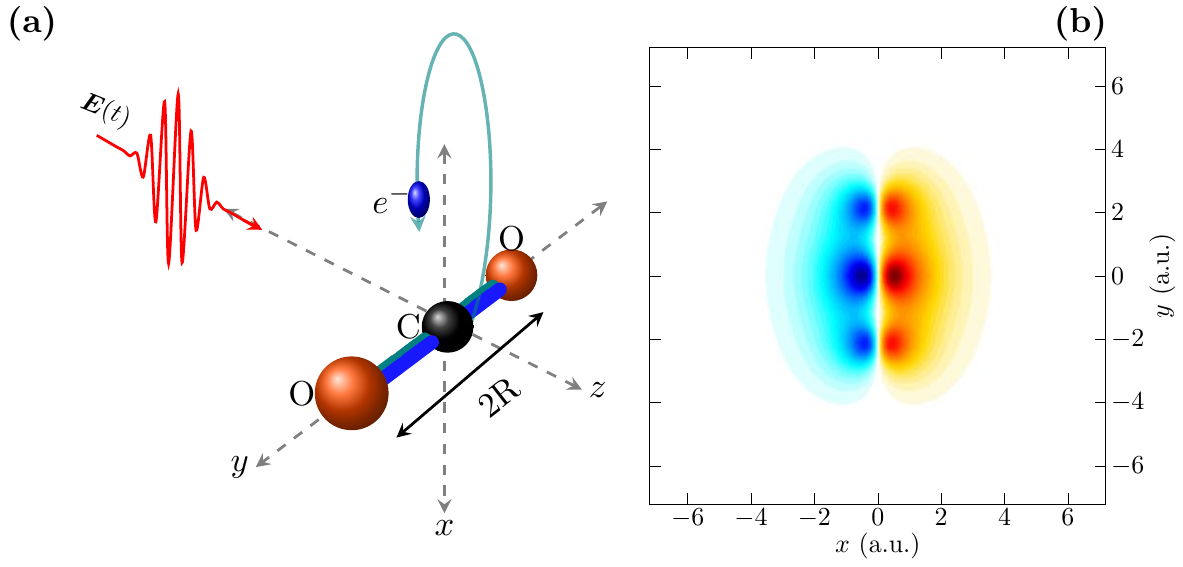}
	\caption{(Color online).  (a) Geometry of the system (see text for details). (b) Schematic representation of the HOMO-1 wave function of a symmetric CO$_2$ molecule with the CO equilibrium internuclear distance $R=R_e\simeq 2.6\,$a.u.}
	\label{fig:system}
\end{figure}

The system considered in Ref.\;\cite{PRA.in.press} is a symmetric, linear, carbon dioxide molecule, CO$_{2}\/$ coupled to an intense IR field. The problem was discussed in detail for the LIED spectra extracted from the HOMO orbital which is mainly localized, in an anti-symmetric manner, on the two oxygen atoms of the molecule\;\cite{PRA.in.press}. Unlike the HOMO, the HOMO-1 of the CO$_{2}\/$ molecule is a symmetric orbital delocalized over all atoms, as shown in Fig.\;\ref{fig:system}(b). Hence during tunnel ionization all atoms of the molecule will play a significant role. This particular structure of the HOMO-1 orbital should imprint its signature in the associated LIED spectra. The HOMO-1 of CO$_{2}\/$ is therefore an ideal candidate for the present work discussing other symmetries as compared to the anti-symmetric HOMO.

The geometry of our system is depicted schematically in Fig.\;\ref{fig:system}(a). The strong field electronic dynamics is triggered by an intense IR laser field. The molecule with symmetric C-O bond length $R$ is assumed to be pre-aligned along the $y$-axis, normal to both the polarization ($x$-axis) and propagation ($z$-axis) directions of the applied intense IR field. This arrangement will mainly confine tunnel ionization and the associated electron dynamics in the plane $xy$ defined by the field polarization and the molecular alignment axes. This 2D dynamics simplifies our problem significantly. The dynamics of the system can then be simulated using an appropriate method, as explained hereafter.   Other cases where the applied field is acting at an angle $ \theta \in \{0,\pi/2\} $ with respect to the molecular axis will be discussed in our following paper of this issue\:\cite{MolPhys.in.press}. This additional paper discusses the robustness of the LIED spectra with respect to the initial alignment of the molecule. Note that, as long as the field is resolved such that the molecular axis and polarization directions are coplanar vectors, the field driven electron dynamics will always be confined into that plane.

\subsection{Numerical Solution of the Time-Dependent Schr\"odinger Equation}\label{subsec:numeric}

The dynamics of the system obeys the time-dependent Schr\"odinger equation (TDSE) given (in atomic units) by
\begin{equation}
\hat{\mathcal{H}}(t)\,|\Psi(t)\rangle = i\,\partial_{t}\,|\Psi(t)\rangle\,,
\label{Eq:TDSE}
\end{equation}
where $|\Psi(t)\rangle$ is the state of the system coupled to the field and $\hat{\mathcal{H}}(t)$ stands for the total Hamiltonian
\begin{equation}
\hat{\mathcal{H}}(t) =  \hat{\mathcal{H}}_{0} - \bm{\mu}\cdot\bm{E}(t)\,,
\label{Eq:Hamiltonian_1}
\end{equation}
where
\begin{equation}
\hat{\mathcal{H}}_{0}=-\bm{\nabla}^2/2+V(\bm{r},\bm{R})
\end{equation}
is the field-free Hamiltonian and $ - \bm{\mu}\cdot\bm{E}(t) $ is the interaction potential taken in the length gauge. The ultra-short linearly polarized interacting laser field $\bm{E}(t)$ is defined via
\begin{equation}
\bm{E}(t)=- \partial_t\,\bm{A}(t)\,,
\end{equation}
where
\begin{equation}
\bm{A}(t) = \frac{E_0}{\omega_L}\,\sin^2\left(\frac{\pi t}{2\tau}\right)\,\cos(\omega_L t + \phi)\,\hat{\bm{e}}_x
\label{Eq:Vect_pot}
\end{equation}
is the vector potential assuming a polarization along the $x$ direction. $\omega_L = 2\pi c/\lambda_L$ is the IR carrier frequency and $E_0$ is the electric field amplitude. $\phi$ is the Carrier-Envelop Phase (CEP) and $\tau$ is the Full Width at Half Maximum (FWHM) of the sine-squared pulse envelop. The binding potential $V(\bm{r},\bm{R})$ is taken as an effective soft-Coulomb potential modeled within the single active electron approximation (SAE)\;\cite{RPP.60.389}, as introduced initially in Ref.\;\cite{PRA.85.053417}.

The SAE approximation and the specific geometry we have chosen (see Fig.\;\ref{fig:system}(a)) reduce the dynamics to a   2D problem, the transverse dynamics along $z$ being limited to a simple spreading effect. The 2D dynamics is solved numerically using the second-order split-operator method\;\cite{JCompP.47.412} on a spatial grid with $\Delta x = \Delta y \simeq 0.5$\,a.u. and 1024 grid points in each direction. The initial state is calculated using the imaginary time propagation (ITP) technique\;\cite{JCompP.221.148}. If the time step $ \Delta t$ used for the ITP is too large, it leads to inaccuracies in the calculated state. In order to avoid such problems we use the following strategy: Once convergence in energy is achieved for a given $\Delta t$, the wave function is further propagated using ITP with a smaller time step. This procedure is repeated until the wave function is relaxed to a state which is unaffected by any further decrease in $\Delta t$. Once this initial state is calculated, the dynamics is solved by propagating the wave function in real time using the method discussed by S. Chelkowski and A. D. Bandrauk in Ref.\;\cite{IJQC.60.1685}.

In principle, the state of the system defined by $|\Psi(t)\rangle$ should describe both the electronic and nuclear dynamics. In order to simplify our model, we neglect the slower nuclear dynamics in comparison with the very fast electron dynamics. This is physically relevant if the system is exposed to an {\em ultra-short} laser pulse within which the nuclear dynamics is negligible. We use here 3.5-optical-cycle pulses, with $\lambda_L = 2.0\,\mu$m. The laser intensity is $10^{14}$\,W/cm$^2$.

\begin{figure}[ht]
	\centering
	\includegraphics[width=0.95\textwidth]{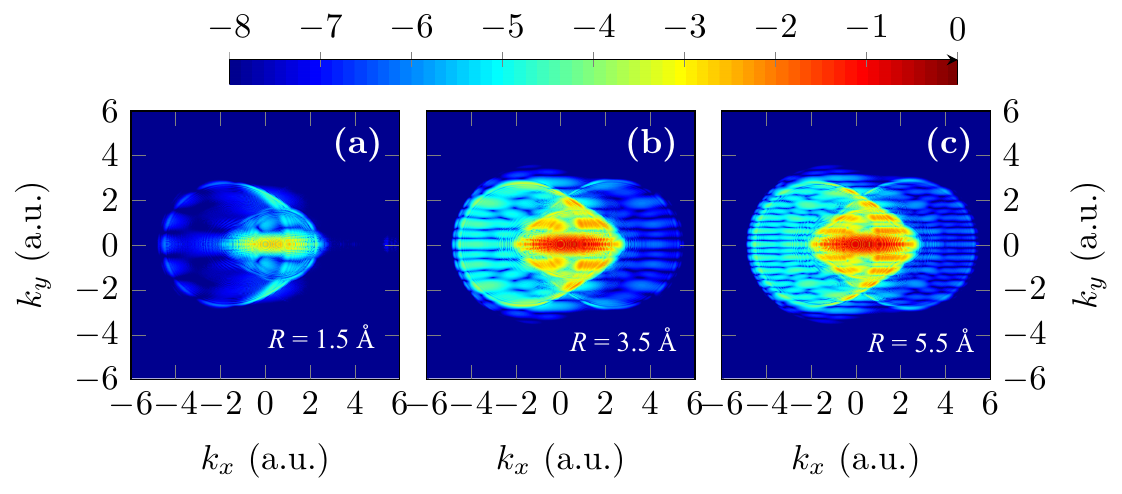}
	\caption{(Color online).  Normalized 2D photoelectron spectra $\mathcal{I}(k_x,k_y)$ in log scale (see the color map) for the HOMO-1 of a symmetric CO$_{2}\/$ molecule obtained for a $3.5$-optical-cycle pulse of intensity $10^{14}$\,W/cm$^2$, wavelength $\lambda_L = 2.0\,\mu$m and CEP $\phi=0$. The panels (a), (b) and (c) are for $R=1.5$\,\AA\/, 3.5\,\AA\/ and 5.5\,\AA\/, respectively.}
	\label{fig:2D_ph_spectra}
\end{figure}

The 2D photoelectron LIED spectrum $\mathcal{I}(k_x,k_y)$ is calculated by projecting the asymptotic part of the electron wave function at the end of the pulse $|\Psi_{\mathrm{as}}(t_f)\rangle$ on a set of 2D plane waves $|\Phi^{\mathrm{pw}}_{\bm{k}}\rangle$ following the procedure developed in Refs.\,\cite{IJQC.60.1685,PRA.52.1450}, as
\begin{equation}
\mathcal{I}(k_x,k_y) = \big|\big\langle\Phi^{\mathrm{pw}}_{\bm{k}}\big|\Psi_{\mathrm{as}}(t_f)\big\rangle \big|^2\,.
\end{equation}

Fig.\;\ref{fig:2D_ph_spectra} shows such a typical 2D photoelectron spectrum obtained from the HOMO-1 orbital of a symmetric CO$_{2}\/$ molecule. Panel (a) is for an internuclear distance $R$ of 1.5\,\AA\/, panel (b) is for $R=3.5$\,\AA\/ and panel (c) shows the same for $R=5.5$\,\AA\/. The key features of typical LIED spectra from the HOMO have been discussed already in Ref.\;\cite{PRA.in.press}. The main difference between the LIED spectra shown in Fig.\;\ref{fig:2D_ph_spectra} and the one for the HOMO is the variation in intensities between successive interference fringes\;\cite{PRA.83.051403,PRA.85.053417}. Another notable difference is the fact that the spectra are characterized by a maximum for the momentum $k_y=0$ ($y$ being the direction of the internuclear axis), whereas this point is a zero (it corresponds to a nodal plane) of the spectrum associated with the HOMO.

It has been shown that it is much easier to study averaged 1D-photoelectron spectra $\mathcal{S}(k_y)$ obtained by averaging out $\mathcal{I}(k_x,k_y)$ over the momentum $k_x$ parallel to the polarization axis\;\cite{PRA.83.051403,PRA.85.053417,PRA.in.press}, using
\begin{equation}
\mathcal{S}(k_y) = \int \mathcal{I}(k_x,k_y) \, dk_x\,.
\label{Eq:1D_spec_Eqn}
\end{equation}
The most interesting feature of these spectra lie in the interference patterns along this parallel momentum axis $k_y$. It was shown in Ref.\,\cite{PRA.83.051403} that the fringe width associated with these structures can be related, in the case of the HOMO, to the internuclear distance $R$ via the relation
\begin{equation}
\Delta k_y=\pi/R\,.
\end{equation}

\begin{figure}[ht]
	\centering
	\includegraphics[width=0.95\textwidth]{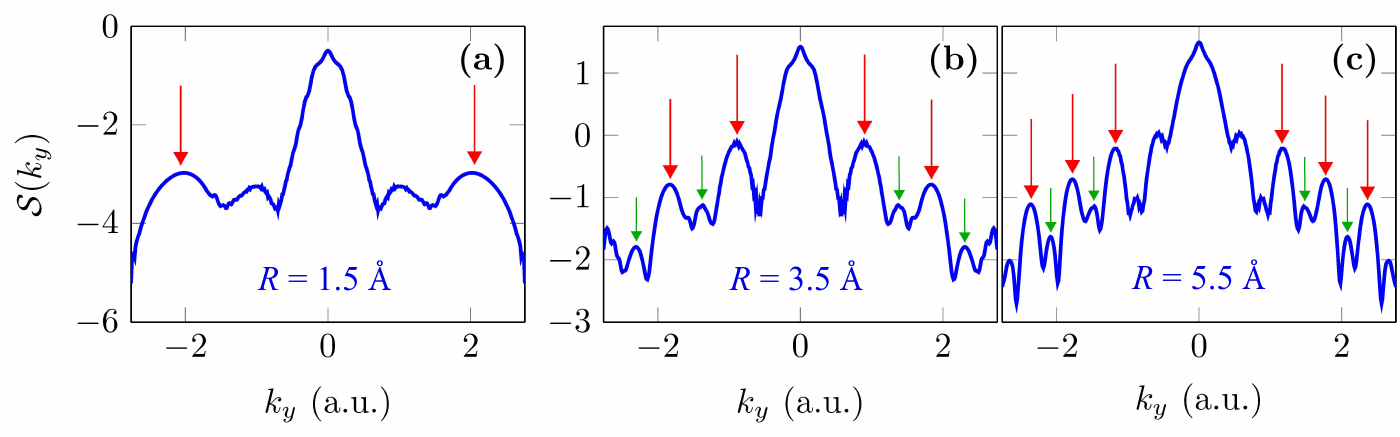}
	\caption{(Color online). 1D averaged photoelectron spectra $\mathcal{S}(k_y)$ associated with the 2D spectra shown in Fig.\;\ref{fig:2D_ph_spectra}, in log scale. The panels (a), (b) and (c) are for $R=1.5$\,\AA\/, $R=3.5$\,\AA\/ and $R=5.5$\,\AA\/, respectively. All other parameters are as in Fig.\;\ref{fig:2D_ph_spectra}.}
	\label{fig:1D_ph_spectra}
\end{figure}

Fig.\;\ref{fig:1D_ph_spectra} shows such 1D-spectra in log scale obtained from averaging the 2D-spectra shown in Fig.\;\ref{fig:2D_ph_spectra}. Each of the   panels labeled (a), (b) and (c) corresponds  to the  one  of Fig.\;\ref{fig:2D_ph_spectra} bearing the same label. In contrast with the simple spectra obtained in the case of the HOMO\;\cite{PRA.in.press}, the 1D averaged photoelectron spectra from the HOMO-1 are characterized by a regular succession of two peaks with different amplitudes\;\cite{PRA.83.051403, PRA.85.053417}. These peaks are marked with long-red and short-green arrows in Fig.\;\ref{fig:1D_ph_spectra}, and these different successive peaks are characterized by the period
\begin{equation}
\Delta k'_y=2\pi/R\,.
\end{equation}

The peaks in the spectra are due to the interference of electron wave packets ionized and scattered from the three different ionic centers. These spectra, hence, imprint all essential information for imaging the molecular orbital from which these electrons were ionized. These interference patterns can resist misalignment defects. This case is discussed in detail in Ref\:\cite{MolPhys.in.press}.  Reading this information is not easy since it is encoded in the spectra in a complicated way. It thus requires the development of a model describing the electronic dynamics of the system such that it can potentially reproduce all the features of the spectra. We will discuss such a simple analytical model in the following subsection.

\subsection{Simplified analytical model based on SFA}\label{subsec:analytic}

We show here how to invert photoelectron spectrum data to obtain an image of the initial state of the system. The photoelectron spectrum is here a calculated one, but the inversion procedure is intended for use on an experimental LIED spectrum leading to the full reconstruction of a molecular orbital of a triatomic, symmetric molecule such as CO$_2$. Such an inversion procedure was proposed in Ref.\;\cite{PRA.in.press} in the case of the HOMO. Here we recall the particular points of this model that are essential for understanding our forthcoming discussion concerning the HOMO-1. For a detailed description of the model, see Ref.\;\cite{PRA.in.press}.

The photoelectron spectrum is calculated by splitting $|\psi(t)\rangle$ in three parts: (i) the part of the wave packet that is still bound to the system (ii) the part describing direct ionization, and (iii) the part describing the recollision process. This separation can be done using the Dyson equation which gives a time-ordered series expansion for the quantum evolution operator\;\cite{PRSA.463.2195, ATI_top_rev, PhysRevA.74.063404, Reiss_SFA}. In general, this series can be continued in order to treat all recollision orders. But it is also possible to concentrate on the first recollision event by approximating the total evolution operator using the Strong Field Approximation (SFA)\;\cite{PhysRevA.74.063404, Reiss_SFA, PhysRevA.78.033412}. As a consequence of this Dyson splitting, the approximate analytical photoelectron spectrum can be written as
\begin{equation}
\mathcal{I}_{\mathrm{ap}}(k_x,k_y) = \big|a_d(k_x,k_y) + a_r(k_x,k_y)\big|^2\,,
\label{Eq:Trans_prob_1}
\end{equation}
where $a_d(k_x,k_y)$ and $a_r(k_x,k_y)$ represent ionization amplitudes associated with a direct process or a recolliding process\;\cite{PRA.in.press}. Eq.(\ref{Eq:Trans_prob_1}) shows that in general these two processes interfere for a given asymptotic electron momentum $(k_x,k_y)$.

It has been shown that large electron energies (and thus small de Broglie wavelengths able to resolve sub-\AA\/ spatial scales) are reached by electrons ionized around a maximum of the field\;\cite{PhysRevA.54.742}. Assuming that the initial ionization process occurs at an instant of time close to this maximum of the field, we obtain
\begin{equation}
a_d(k_x,k_y) = A_d\;\big\langle\Phi^{\mathrm{pw}}_{\bm{k}}\big|x\big|\Psi(0)\big\rangle
\label{Eq:direct_amp_SFA1}
\end{equation}
and
\begin{equation}
a_r(k_x,k_y) = A_d\int\!\!d\bm{k'}\,e^{-ik'^2\Delta t/2}\,\big\langle\Phi^{\mathrm{pw}}_{\bm{k}}\big|V(\bm{r},\bm{R})\big|\Phi^{\mathrm{pw}}_{\bm{k'}}\big\rangle
                                                          \big\langle\Phi^{\mathrm{pw}}_{\bm{k'}}\big|x\big|\Psi(0)\big\rangle\,.
\label{Eq:Recollision_amp_SFA1}
\end{equation}
where $\Delta t$ denotes the average time for the classical excursion of ionized wave packets in the continuum\;\cite{PhysRevA.54.742}. $A_{d}$ and $A_{r}$ represent scaling parameters that will be used in the inversion procedure described in the next section.

In order to evaluate the ionization amplitudes $a_d(k_x,k_y)$ and $a_r(k_x,k_y)$, and therefore the photoelectron spectrum $\mathcal{I}(k_x,k_y)$, we   represent the initial molecular wave function in the form of a simple linear combination of $2p_x$ atomic Gaussian-type orbitals (GTO) given by
\begin{equation}
\Phi_{2p_x}(x,y) = \mathcal{N}_g\;x\;\mathrm{e}^{-\alpha\,(x^2+y^2)}
\label{Eq:STO-1G-2px}
\end{equation}
where $\mathcal{N}_g = \alpha\,\sqrt{8/\pi}$ is the normalization constant in 2D and $\alpha$ is the Gaussian exponent. This choice simplifies significantly the evaluation of the integrals (\ref{Eq:direct_amp_SFA1}) and (\ref{Eq:Recollision_amp_SFA1}). For the HOMO-1 of $CO_2$, the initial molecular wave function is taken as
\begin{equation}
\big|\Psi(0)\big\rangle =   \xi_{\mathrm{o}}\,|\Phi_{\mathrm{2p}_x}^{-}\rangle
                          +                   |\Phi_{\mathrm{2p}_x}\rangle
                          + \xi_{\mathrm{o}}\,|\Phi_{\mathrm{2p}_x}^{+}\rangle\,,
\label{Eq:mol_orb_general}
\end{equation}
where $\xi_{\mathrm{o}}$ is the relative weighting factor for the oxygen atoms with respect to the carbon atom, and 
\begin{eqnarray}
|\Phi_{\mathrm{2p}_x}^{\pm}\rangle = \hat{\mathcal{T}}_{\pm \bm{R}}\,|\Phi_{\mathrm{2p}_x}\rangle\,,
\end{eqnarray}
$\hat{\mathcal{T}}_{\pm\bm{R}}$ denoting the translation operator by the displacement vector $\pm\bm{R}$. 
With these   assumptions, we finally obtain
\begin{equation}
a_{d}(k_x,k_y) = A_{d}\,(k_x^2-2\alpha)\,\mathrm{e}^{-\frac{k_x^2 +k_y^2}{4\alpha}}\,\big[2\xi_{\mathrm{o}}\cos(k_y R)+1\big]\,.
\label{Eq:HOMO-1_direct_CO2}
\end{equation}
for the direct ionization amplitude.

\begin{figure}[ht]
	\centering
	\includegraphics[width=0.3\textwidth]{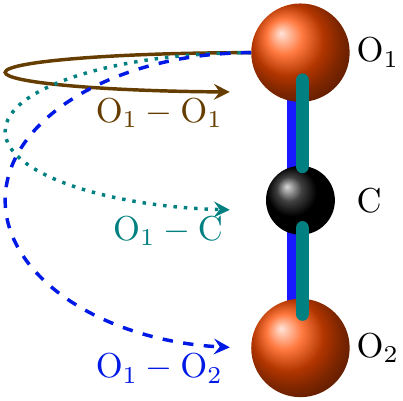}
	\caption{(Color online). Schematic representation of recollision pathways of wave packets ionized from one of the oxygen atoms labeled O$_1$. The three generated recollision pathways (O$_1$-O$_1$, O$_1$-C and O$_1$-O$_2$) are illustrated by the three different arrows (see text for details).}
	\label{Fig:Recollision_scheme}
\end{figure}

The calculation of the recolliding amplitude $a_{r}(k_x,k_y)$ given in Eq.\;(\ref{Eq:Recollision_amp_SFA1}) is more involved. Indeed, the initial wave function (\ref{Eq:mol_orb_general}) is written as a sum of three different atomic orbitals. In addition, the binding potential $V(\bm{r},\bm{R})$ seen in Eq.\;(\ref{Eq:Recollision_amp_SFA1}) is characterized by three attraction centers. As a consequence, the evaluation of $a_{r}(k_x,k_y)$ requires the calculation of nine different integrals. This is because in the HOMO-1 the three atoms of the molecule serve both as ionization sources and as rescattering centers. Fortunately, these nine different integrals are similar. Each integral corresponds to a specific ionization-recollision pathway. Three of these nine different pathways are illustrated in Fig.\;\ref{Fig:Recollision_scheme}. They correspond to the ionization from one of the atoms: here the oxygen atom labeled as O$_1$. On recollision, this contribution from O$_1$ will scatter from the parent atom O$_1$ itself (path shown as a solid brown arrow labeled O$_1$-O$_1$ in this figure) and also from its two neighbors: from the carbon atom (shown as a dotted green arrow labeled O$_1$-C) as well as from the second oxygen atom (shown as a dashed blue arrow labeled O$_1$-O$_2$).

Finally, the evaluation of these nine integrals yields the following approximate recollision amplitude
\begin{equation}
a_{r}(k_x,k_y) = \frac{A_{r}}{\left| k_y \right|}\,\Big[
\cos(k_y R)\left(\mathrm{e}^{i\beta R^2}+\xi_{\mathrm{o}}\,\mathrm{e}^{i 4\beta R^2}+\xi_{\mathrm{o}}\right)
+ \frac{1}{2} + \xi_{\mathrm{o}}\,\mathrm{e}^{i\beta R^2}\Big]\,,
\label{Eq:HOMO-1_recoll_CO2}
\end{equation}
where $\beta^{-1}=2\Delta t$. Summing the amplitudes given by Eqs.\;(\ref{Eq:HOMO-1_direct_CO2}) and\;(\ref{Eq:HOMO-1_recoll_CO2}) gives access to the 2D approximate photoelectron spectrum $\mathcal{I}_{\mathrm{ap}}(k_x,k_y)$ of Eq.\;(\ref{Eq:Trans_prob_1}). Averaging finally this 2D spectrum along $k_x$ gives the 1D approximate LIED spectrum
\begin{eqnarray}
\mathcal{S}_{\mathrm{ap}}(k_y) & = &  A_d^2\,\mathrm{e}^{-\frac{k_y^2}{2 \alpha }}\,\big[2\xi_{\mathrm{o}}\cos(k_y R)+1\big]^2 + \nonumber \\[0.2cm]
                              & + & \frac{A_r^2}{k_y^2}\,\Big|\cos(k_y R)\left(\mathrm{e}^{i\beta R^2}+
                                    \xi_{\mathrm{o}}\,\mathrm{e}^{i 4\beta R^2}+\xi_{\mathrm{o}}\right) +
                                    \frac{1}{2} + \xi_{\mathrm{o}}\,\mathrm{e}^{i\beta R^2}\Big|^2\,.
\label{Eq:1D_spectrum_HOMO-1}
\end{eqnarray}
This model will be used in the next section to extract information about the initial molecular orbital from the spectra $\mathcal{S}(k_y)$ calculated from the time-dependent Schr\"odinger equation.

\section{Reconstruction of the HOMO-1 orbital}\label{sec:reconst}

The analytical formula we have obtained in Eq.\;(\ref{Eq:1D_spectrum_HOMO-1}) is based on an approximate model. It can be compared with the `exact' 1D spectra shown in Fig.\;\ref{fig:1D_ph_spectra}. To reproduce the exact spectra from the analytical model there are five parameters that have to be optimized with a fitting procedure: $A_r$, $A_d$, $\alpha$, $\xi_{\mathrm{o}}$ and $R$. We use the well-known Levenberg-Marquardt algorithm (LMA)\;\cite{Levenberg, Marquardt} for fitting our model.

While deriving the simple form given in Eq.\;(\ref{Eq:1D_spectrum_HOMO-1}) we have made several assumptions and one has to keep those assumptions in mind. The most important one is that the model is made to describe the high-energy part of the spectra only. Thus, while fitting the $k_x$-averaged photoelectron spectra, one should aim at getting the best fit for the fastest electrons. In practice we perform the fitting procedure within a particular range $[k_y^{\mathrm{min}},k_y^{\mathrm{max}}]$. The fitting procedure can be repeated by changing this range of momentum. Since the most interesting part for the model where it holds relatively well is the highest momentum part, the upper limit of the fitting range is fixed using the cut-off energy $3.17\,U_p$, where $U_p$ is the ponderomotive energy\;\cite{RMP.81.163}. Now, by changing the lower limit $k_y^{\mathrm{min}}$ we can repeat the fitting procedure. Table\;\ref{table:fitted_values} shows a series of parameters retrieved by fitting the model to exact spectra for different values of $k_y^{\mathrm{min}}$. The calculations are done for three internuclear distances: $R=1.5$\,\AA\/, 3.5\,\AA\/ and 5.5\,\AA.

\begin{figure}[ht]
	\centering
	\includegraphics[width=0.95\textwidth]{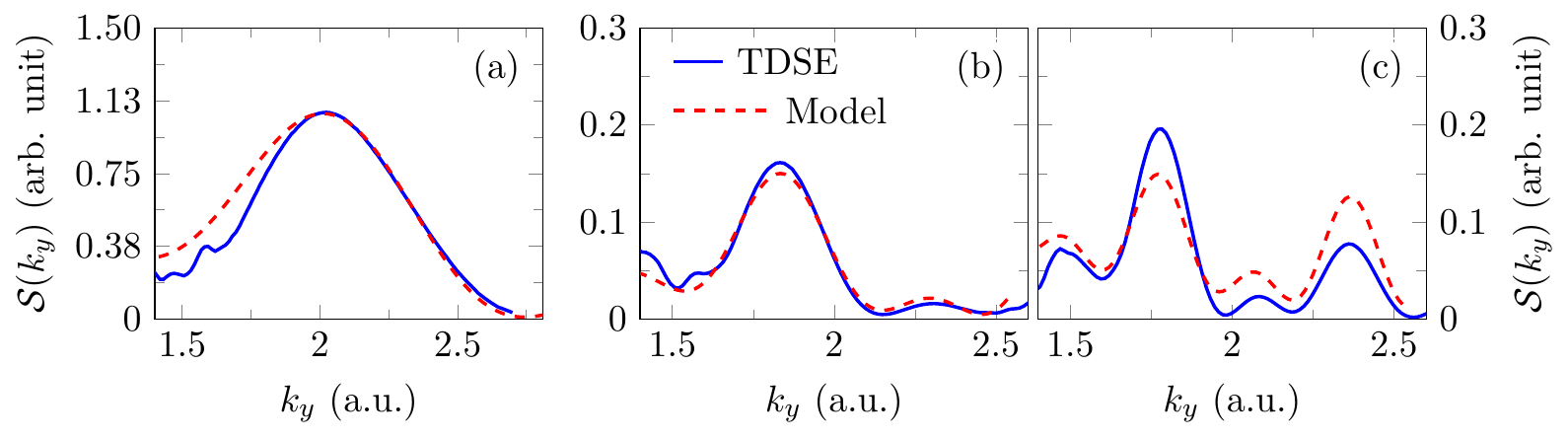}
	\caption{(Color online). Comparison of calculated and fitted photo-electron spectra for different internuclear distances. (a) is for $R=1.5$\,\AA\/, (b) is for $R=3.5$\,\AA\/ and (c) is for $R=5.5$\,\AA\/. All other parameters are as in Fig.\;\ref{fig:2D_ph_spectra}. The blue curves are the calculated spectra and the red dashed curves are the fitted spectra using Eq.\;(\ref{Eq:1D_spectrum_HOMO-1}).}
	\label{Fig:fitted_ph_spec}
\end{figure}

To reconstruct the initial state, the average values of the parameters given in Table\;\ref{table:fitted_values} will be taken. Fig.\;\ref{Fig:fitted_ph_spec} shows the comparison between the exact and the fitted 1D photoelectron spectra. The blue solid curves are the exact spectra from the numerical solution of the TDSE and the red dashed curves are the model. Panel (a) shows the comparison for $R=1.5$\,\AA\/ and panels (b) and (c)  show the same for $R=3.5$\,\AA\/ and $R=5.5$\,\AA\/, respectively. The fitted  spectra agree fairly well with the calculated ones.

\begin{table}[ht]
\tbl{Parameters retrieved from the photoelectron spectra of Fig.\;\ref{fig:1D_ph_spectra} using Eq.\,(\ref{Eq:1D_spectrum_HOMO-1}).}
{\begin{tabular}{cccccc} \toprule
Case 		  & $A_d$ (a.u.)  & $A_r$ (a.u.)   & $\alpha$ (a.u.) &  $\xi_{\mathrm{o}}$ (a.u.)  & $R$ (\AA)    \\[0.5ex]
\hline \noalign{\vskip 1.0ex}
			  & 0.036    	  & 0.016    	   & 0.830           &   0.140  		           & 1.622        \\[0.5ex]
			  & 0.042 		  & 0.014    	   & 0.810           &   0.186  		           & 1.614        \\[0.5ex]
$R=1.5$\,\AA  & 0.045  		  & 0.013          & 0.850           &   0.165                     & 1.598        \\[0.5ex]
			  & 0.049  		  & 0.013          & 0.840           &   0.163 	                   & 1.614        \\[1.5ex]
{\em Average} & 0.043  		  & 0.014          & 0.833           &   0.164  	               & 1.612        \\[0.5ex]
\hline \noalign{\vskip 1.0ex}
			  & 0.339   	  & 0.224    	   & 0.886           &   0.051                     & 3.615        \\[0.5ex]
			  & 0.332  		  & 0.253		   & 0.812           &   0.027                     & 3.612        \\[0.5ex]
$R=3.5$\,\AA  & 0.338   	  & 0.242  		   & 0.855           &   0.032 	                   & 3.615        \\[0.5ex]
			  & 0.333 		  & 0.227    	   & 0.875           &   0.066                     & 3.616        \\[1.5ex]
{\em Average} & 0.335    	  & 0.236   	   & 0.857           &   0.044 	                   & 3.615        \\[0.5ex]
\hline \noalign{\vskip 1.0ex}
			  & 0.428  		  & 0.246          & 0.879           &   $ \approx 0 $             & 5.620        \\[0.5ex]
			  & 0.459    	  & 0.245          & 0.870           &   $ \approx 0 $             & 5.590        \\[0.5ex]
$R=5.5$\,\AA  & 0.457  		  & 0.213          & 0.876           &   $ \approx 0 $             & 5.630        \\[0.5ex]
			  & 0.445   	  & 0.246          & 0.865           &   $ \approx 0 $             & 5.630        \\[1.5ex]
{\em Average} & 0.447    	  & 0.237          & 0.873           &   $ \approx 0 $             & 5.618        \\[0.5ex]
\bottomrule
\end{tabular}}
\label{table:fitted_values}
\end{table}

Note that the model we developed is based on a a representation of  the initial wave function using Gaussian-type orbitals. This practical expression helped us to get a simple analytical form for the LIED spectra. On the other hand, it is well-known that Slater-type orbitals (STO) are more accurate for describing molecular orbitals. Thus we now optimize the reconstructed orbitals using  STOs  defined by 
\begin{equation}
\Phi_{2p_x}(x,y)=\mathcal{N}_s\;x\;\mathrm{e}^{-\zeta\,\sqrt{x^2+y^2}}\,,
\label{Eq:STO_2px}
\end{equation}
where $\zeta$ is the Slater exponent and $\mathcal{N}_s=\zeta^2/\sqrt{8/3\pi}$ is the normalization constant in 2D. It can be shown that the best optimized Slater exponent is given by \mbox{$\zeta=2.165\sqrt{\alpha}$}\;\cite{PRA.in.press}. We now use this particular Slater exponent to represent the reconstructed molecular orbitals.

\begin{figure}[ht]
 	\centering
 	\includegraphics[width=0.9\textwidth]{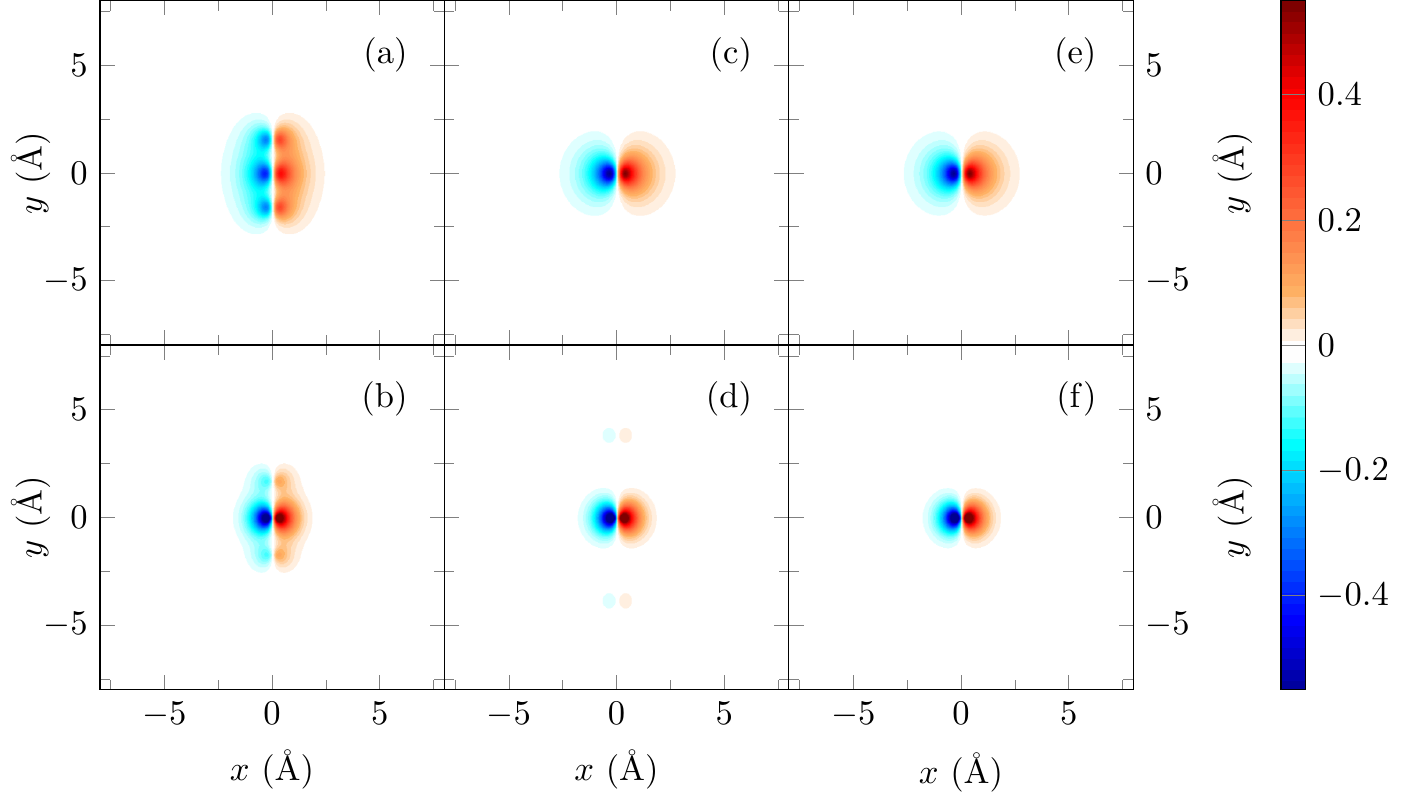}
 	\caption{(Color online). Initial wave functions and their Slater reconstructions. Panels (a), (c) and (e) are the exact initial states for $R = 1.5$\,\AA\/, 3.5\,\AA\/ and 5.5\,\AA\/, respectively. Panels (b), (d) and (f) are the corresponding Slater reconstructions.}
 	\label{Fig:reconstruction}
\end{figure}

\begin{figure}[ht]
 	\centering
 	\includegraphics[width=0.9\textwidth]{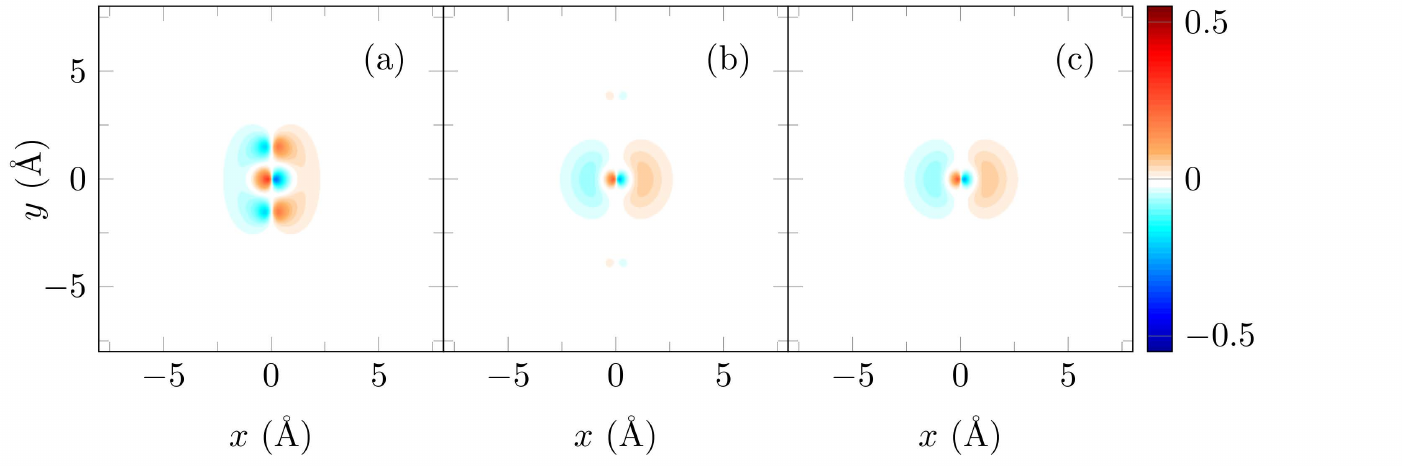}
 	\caption{(Color online). Difference between the initial and reconstructed molecular orbitals. Panel (a) shows the difference between the initial and reconstructed Slater orbitals for $R = 1.5$\,\AA\/. Panels (b) and (c) show the same for $R = 3.5$\,\AA\/ and $R = 5.5$\,\AA\/. The color map used here is the same as in Fig.\;\ref{Fig:reconstruction}.}
 	\label{Fig:diff_states}
\end{figure}

Fig.\;\ref{Fig:reconstruction} shows the exact HOMO-1 orbitals (first row) and their reconstructions using STO representation of the initial wavefunction, (second line) using the data given in Table\;\ref{table:fitted_values}. Panels (a), (c) and (e) are the exact initial states used in the calculation of the LIED spectra for $R = 1.5$\,\AA\/, 3.5\,\AA\/ and 5.5\,\AA\/, respectively. Panels  (b), (d) and (f) are the corresponding Slater reconstructions. The overlap between the initial and reconstructed wave functions for $R = 1.5$\,\AA\/ is 0.87. For $R = 3.5$\,\AA\/ and $R = 5.5$\,\AA\/ the two wave functions have an overlap close to 0.95. For a qualitative comparison, the difference between the initial and the reconstructed wave functions are given in Fig.\;\ref{Fig:diff_states} using the same color code as in Fig.\;\ref{Fig:reconstruction}. As we have already seen from the HOMO discussed in Ref.\;\cite{PRA.in.press}, the reconstruction is of relatively good quality, and the more so, the larger the internuclear distance.

This reconstruction procedure is able to extract an important information from the LIED spectra which consists in the fact that, as the internuclear distance increases, the electron appears more localized on the central carbon atom. Indeed, at $R = 5.5\,$\AA\/, the contribution from the oxygen atoms becomes zero. Thus for this internuclear distance, the ionization signal has no contributions from the oxygen atoms. However, even if they are not contributing to the ionization signal, they play a significant role in the recollision process and the presence of these atoms is imprinted in the LIED spectra as a double period in the interference patterns which would be absent otherwise\;\cite{PRA.83.051403, PRA.85.053417}. Note that in the approximate numerical model used here the binding potential is constituted by a succession of three adjacent wells separated by the internuclear distance $R$. The typical reconstruction example shown in Fig.\;\ref{Fig:reconstruction} proves the ability of elastically recolliding wave packets and hence of LIED to image the structure of molecular orbitals.

From Figs.\;\ref{Fig:reconstruction} and\;\ref{Fig:diff_states} it appears clearly that the model discussed here can be used for any internuclear distance, but due to the natural limit set by the de Broglie wavelength of recolliding electron wave packets, as we go to smaller distances, the spectra will be dominated mainly by low-energy electrons (see Fig.\;\ref{fig:2D_ph_spectra}) where the model does not hold very well. Indeed, the error in the retrieval of molecular parameters is decreasing as $R$ increases.

\section{Conclusions}\label{sec:conclude}

In this paper, we have discussed the possibility of imaging molecular orbitals of linear molecules by using a simple analytical SFA model. The ability of this imaging procedure is demonstrated here for the particular case of the HOMO-1 molecular orbital of the carbon dioxide molecule.

The LIED spectra are calculated by solving the time dependent Schr\"odinger equation in a single-active electron model of the CO$_{2}\/$ molecule. An analytical model based on the Strong Field Approximation is discussed. It consists in an extension of the model we have already introduced in Ref.\;\cite{PRA.in.press}. This model is used to extract a set of parameters relevant for describing the initial state of the system. The internuclear distance is extracted with an error of less than 8\% close to the equilibrium distance. This error becomes much smaller for higher internuclear distances (around 2\% at $R=5.5$\,\AA\/). The initial electronic state is finally reconstructed for different internuclear distances using optimized Slater-type orbitals. The reconstructed orbitals compare well with the initial states used for calculating the photoelectron spectra.

The model was able to reproduce various delocalizations of the initial electronic cloud at different internuclear distances. It shows the potential of such analytical models to describe the dynamics of a dissociative process. In the future, we plan to extend our model for such time-resolved dynamics in order to propose an imaging technique able to capture the nuclear dynamics of molecular processes such as photo-dissociation processes.

\section*{Acknowledgment}

R.P.J. and E.C. acknowledge support from  the EU (Project ITN - 2010 - 264951, CORINF). We thank Misha Ivanov for fruitful discussions. We also acknowledge the use of the computing cluster GMPCS of the LUMAT federation (FR 2764 CNRS).  O.A. acknowledges the organizing committee of the \textit{Andr\'e D. Bandrauk Honorary Symposium on Molecules and Laser Fields} in Orford (QC), Canada, May 2016, for giving him the opportunity of an invited talk partly covering the subject of this article.




\begin{thebibliography}{100}

\bibitem{JETP.20.1307}
L. V. Keldysh.
\newblock {\em Soviet Physics JETP}, \textbf{20}, 1307, (1965).

\bibitem{PRL.71.1994}
P. B. Corkum.
\newblock {\em Phys. Rev. Lett.}, \textbf{71}, 1994, (1993).

\bibitem{CPL.259.313}
T. Zuo, A. D. Bandrauk, and P. B. Corkum.
\newblock {\em Chem. Phys. Lett.}, \textbf{259}, 313, (1996).

\bibitem{Sc.320.1478}
M. Meckel, D. Comtois, D. Zeidler, A. Staudte, D. Pavicic, H. C. Bandulet, H. Pepin, J. C. Kieffer, R. Dorner, D. M. Villeneuve, and P. B. Corkum.
\newblock {\em Science}, \textbf{320}, 1478, (2008).

\bibitem{PRA.83.051403}
M. Peters, T. T. Nguyen-Dang, C. Cornaggia, S. Saugout, E. Charron, A. Keller, and O. Atabek.
\newblock {\em Phys. Rev. A}, \textbf{83}, 051403(R), (2011).

\bibitem{PRA.85.053417}
M. Peters, T. T. Nguyen-Dang, E. Charron, A. Keller, and O. Atabek.
\newblock {\em Phys. Rev. A}, \textbf{85}, 053417, (2012).

\bibitem{Nat.483.194}
C. I. Blaga, J. Xu, A. D. DiChiara, E. Sistrunk, K. Zhang, P. Agostini, T. A. Miller, L. F. DiMauro, and C. D. Lin.
\newblock {\em Nature}, \textbf{483}, 194, (2012).

\bibitem{NatCom.6.7262}
M. G. Pullen, B. Wolter, A.-T. Le, M. Baudisch, M. Hemmer, A. Senftleben, C.-D. Schroter, J. Ullrich, R. Moshammer, C. D. Lin, and J. Biegert.
\newblock {\em Nat. Commun.}, \textbf{6}, 7262, (2015).

\bibitem{JPB.49.112001}
J. Xu, C. I Blaga, P. Agostini, and L. F. DiMauro.
\newblock {\em J. Phys. B: At. Mol. Opt. Phys.}, \textbf{49}, 112001, (2016).

\bibitem{PRL.108.263003}
\newblock X. B. Bian and A. D. Bandrauk.
\newblock {\em Phys. Rev. Lett.}, \textbf{108}, 263003, (2012).

\bibitem{PRA.84.043420}
\newblock X. B. Bian, Y. Huismans, O. Smirnova, K. J. Yuan, M. J. J. Vrakking, and A. D. Bandrauk.
\newblock {\em Phys. Rev. A}, \textbf{84}, 043420, (2011).

\bibitem{Nat.432.867}
J. Itatani, J. Levesque, D. Zeidler, H. Niikura, H. Pepin, J. C. Kieffer, P. B. Corkum, and D. M. Villeneuve.
\newblock {\em Nature}, \textbf{432}, 867, (2004).

\bibitem{Nat.Phys.7.822}
C. Vozzi, M. Negro, F. Calegari, G. Sansone, M. Nisoli, S. De Silvestri, and S. Stagira.
\newblock {\em Nat. Phys.}, \textbf{7}, 822, (2011).

\bibitem{PRA.in.press}
R. Puthumpally-Joseph, J. Viau-Trudel, M. Peters, T. T. Nguyen-Dang, O. Atabek, and E. Charron.
\newblock {\em Phys. Rev. A}, \textbf{94}, 023421, (2016).

\bibitem{MolPhys.in.press}
T. T. Nguyen-Dang, M. Peters, J. Viau-Trudel, E. Couture-Bienvenue, R. Puthumpally-Joseph, E. Charron and O. Atabek.
\newblock {\em Mol. Phys (Submitted)}, (2016).

\bibitem{RPP.60.389}
M. Protopapas, C. H. Keitel, and P. L. Knight.
\newblock {\em Rep. Prog. Phys.}, \textbf{60}, 389, (1997).

\bibitem{JCompP.47.412}
M. J. Feit, J. A. Fleck, and A. Steiger.
\newblock {\em J. Comput. Phys.} \textbf{47}, 412, (1982).

\bibitem{JCompP.221.148}
L. Lehtovaara, J. Toivanen, and J. Eloranta.
\newblock {\em J. Comput. Phys.}, \textbf{221}, 148, (2007).

\bibitem{IJQC.60.1685}
S. Chelkowski and A. D. Bandrauk.
\newblock {\em Int. J. Quant. Chem.}, \textbf{60}, 1685, (1996).

\bibitem{PRA.52.1450}
A. Keller.
\newblock {\em Phys. Rev. A}, \textbf{52}, 1450, (1995).

\bibitem{PRSA.463.2195}
M. Frasca.
\newblock {\em Proc. R. Soc. A}, \textbf{463}, 2195, 2007.

\bibitem{ATI_top_rev}
\newblock Above-threshold ionization by few-cycle pulses.
\newblock {\em J. Phys. B: At. Mol. Opt. Phys.}, \textbf{39}, R203, (2006).

\bibitem{PhysRevA.74.063404}
D. B. Milosevic.
\newblock {\em Phys. Rev. A}, \textbf{74}, 063404, (2006).

\bibitem{Reiss_SFA}
H. R. Reiss.
\newblock Foundations of the strong-field approximation.
\newblock {\em Progress in Ultrafast Intense Laser Science III}, volume 89 of {\em Springer Series in Chemical Physics}, page 1.
Springer Berlin Heidelberg, (2008).

\bibitem{PhysRevA.78.033412}
M. Busuladzic, A. Gazibegovic-Busuladzic, D. B. Milosevic, and W. Becker.
\newblock {\em Phys. Rev. A}, \textbf{78}, 033412, (2008).

\bibitem{PhysRevA.54.742}
M. Y. Ivanov, T. Brabec, and N. Burnett.
\newblock {\em Phys. Rev. A}, \textbf{54}, 742, (1996).

\bibitem{Levenberg}
K. Levenberg.
\newblock {\em Quart. Appl. Math.} \textbf{2}, 164, (1944).

\bibitem{Marquardt}
D. Marquardt.
\newblock {\em SIAM J. Appl. Math.} \textbf{11}, 431, (1963).

\bibitem{RMP.81.163}
F. Krausz and M. Ivanov.
\newblock {\em Rev. Mod. Phys.} \textbf{81}, 163, (2009).

\end{thebibliography}
\end{document}